\documentclass[prl,twocolumn,aps,showpacs]{revtex4}

\usepackage{graphicx}
\begin{document}
\title{Comment on paper by L. M. Malyshkin and S. Boldyrev, ''Magnetic dynamo action at low magnetic Prandtl numbers'', PRL 105, 215002 (2010)}
\author{Nathan Kleeorin}
\author{Igor Rogachevskii}
\affiliation{Department of Mechanical Engineering, The Ben-Gurion
University of the Negev, POB 653, Beer-Sheva 84105, Israel}

\date{\today}
\begin{abstract}
Is the scaling, $\lambda \propto {\it Rm}^{1/2}$, for the growth rate of small-scale dynamo instability at low magnetic Prandtl numbers and large magnetic Reynolds numbers, ${\it Rm}$, valid in the vicinity of the threshold? Our analysis and even numerical solution \cite{MB10} of the dynamo equations for a Gaussian white-noise velocity field (the Kazantsev-Kraichnan model) imply that the answer is negative. Contrary to the claim in \cite{MB10}, there are two different asymptotics for the dynamo growth rate: in the vicinity of the threshold and far from the threshold.
\end{abstract}


\maketitle

Let us discuss the asymptotic behaviour of the growth rate of magnetic fluctuations with a zero mean field for small magnetic Prandtl numbers in a homogeneous, isotropic, non-helical, incompressible and Gaussian white-noise velocity field (the Kazantsev-Kraichnan model). The equation for the longitudinal correlation function, $W(r) = \langle b_{r}({\bf x}) \, b_{r}({\bf y}) \rangle$ of the magnetic field reads:
\begin{eqnarray}
18 \, r^2 \, W'' + 96 \, r \, W' + \left(104 - 27 \, \lambda \, r^{2/3} \right) \, W = 0 \,,
\label{B3}
\end{eqnarray}
(see  \cite{AH07}), where $b_r$ is the component of magnetic field ${\bf b}$ in the direction ${\bf r} = {\bf x} - {\bf y}$, $\, W' = dW(r) / dr$, $\, \lambda$ is the growth rate of small-scale dynamo instability, and velocity fluctuations have Kolmogorov scaling from viscous scale to integral scale. Equation~(\ref{B3}) is written in dimensionless
variables: length and velocity are measured in units of $\ell_0$ and
$u_0$, where $u_0$ is the characteristic turbulent velocity in the integral scale $\ell_0$. The solution of Eq.~(\ref{B3}) is $W(r) = C \, r^{-13/6} K_{\alpha} \left(\sqrt{27 \,\lambda/2} \, \, r^{1/3} \right)$ (see  \cite{AH07}), where $K_{\alpha}(y)$ is the real part of the modified Bessel function (Macdonald function) with $\alpha = (i/2) \, \sqrt{39}$. This solution is chosen to be finite at large $r$, with positively defined spectrum, and it has the following asymptotics:
$W(r) = A_1 \, r^{-13/6} \, \cos\left(\ln r + \varphi_0 \right)$ at scales $\lambda^{1/2} \, r^{1/3} \ll 1$ (see \cite{RK97}), and $W(r) = A_2 \, r^{-7/3} \exp \left(-\sqrt{27 \,
\lambda/2} \, \, r^{1/3} \right)$ at scales $\lambda^{1/2} \, r^{1/3} \gg 1$ (see \cite{BC04}).

For $\ell \ge \ell_\eta$ the scaling for the growth rate of small-scale dynamo instability which is far from the threshold, is $\lambda \sim u_\eta /\ell_\eta \sim (u_0 /\ell_0) \, {\it Rm}^{1/2}$ (see \cite{M61}), where $\ell_\eta = \ell_0/{\it Rm}^{3/4}$ is the resistive scale, $u_\eta= (\varepsilon \, \ell_\eta)^{1/3}$ is the characteristic turbulent velocity at the  resistive scale, $u_0= (\varepsilon \, \ell_0)^{1/3}$, $\, \varepsilon = u_0^3 /\ell_0$ is the dissipation rate of turbulent kinetic energy, ${\it Rm}=u_0 \, \ell_0/ \eta \gg 1$ is the magnetic Reynolds number and $\eta$ is the magnetic diffusion due to electrical conductivity of the fluid. For the scaling $\lambda \propto {\it Rm}^{1/2}$, the condition $\lambda^{1/2} \, r^{1/3} \gg 1$ implies $r \gg {\it Rm}^{-3/4}$.

However, the scaling, $\lambda \propto {\it Rm}^{1/2}$, is not valid in the vicinity of the threshold of the dynamo instability.
Indeed, in the vicinity of the threshold when $\lambda \to 0$, there is only one range of the solution of Eq.~(\ref{B3}), $\lambda^{1/2} \, r^{1/3} \ll 1$, which determines the growth rate of the small-scale dynamo instability, $\lambda=\beta \, \ln\left({\it Rm} / {\it Rm}_{\rm cr}\right)$ (see  \cite{RK97}), where $\beta=4/3$ is the exponent of the turbulent diffusivity scaling, $D(\ell) \propto \ell^\beta$. In Fig.~1 we plot the growth rate of small-scale dynamo instability versus $\ln\left({\it Rm}/{\it Rm}_{\rm cr}\right)$ in the vicinity of the threshold, which demonstrates perfect agreement between the scaling $\lambda=\beta \, \ln\left({\it Rm} / {\it Rm}_{\rm cr}\right)$ (solid line) and the numerical solution \cite{MB10} of the dynamo equations for the Kazantsev-Kraichnan model (squares).

\begin{figure}
\centering
\includegraphics[width=7cm]{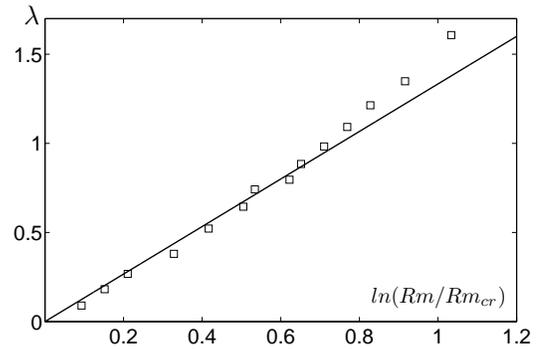}
\caption{\label{Fig1} The growth rate of small-scale dynamo instability versus $\ln\left({\it Rm}/{\it Rm}_{\rm cr}\right)$ in the vicinity of the instability threshold: solid line corresponds to the scaling $\lambda=\beta \, \ln\left({\it Rm} / {\it Rm}_{\rm cr}\right)$ and squares are the results of the numerical solution of the dynamo equations for the Kazantsev-Kraichnan model of velocity field with zero kinetic helicity taken from Fig.~1 in \cite{MB10}.}
\end{figure}

\end{document}